\documentclass[twocolumn,showpacs,preprintnumbers,amsmath,amssymb,superscriptaddress,floatfix]{revtex4}

\usepackage{graphicx}
\usepackage{dcolumn}
\usepackage{bm}
\usepackage{latexsym}
\newcommand{\beqn}{\begin{eqnarray}}
\newcommand{\eeqn}{\end{eqnarray}}

\newcommand{\Kanazawa}{\affiliation{Institute for Theoretical Physics,
Kanazawa University, Kanazawa 920-1192, Japan}}
\newcommand{\RIKEN}{\affiliation{RIKEN, Radiation Laboratory, Wako 351-0158, Japan}}
\newcommand{\Mainz}{\affiliation{Numazu College of Technology, Numazu 410-8501, Japan}}

\begin{document}

\preprint{KANAZAWA 07-09}

\title{Gauge-independent Abelian mechanism of color confinement in  gluodynamics
}
\author{Tsuneo Suzuki}
\Kanazawa
\RIKEN
\author{Katsuya Ishiguro}
\Kanazawa
\RIKEN
\author{Yoshiaki Koma}
\Mainz
\author{Toru Sekido}
\Kanazawa
\RIKEN

\date{\today}

\begin{abstract} 
Abelian mechanism of non-Abelian color confinement is 
observed in a gauge-independent way by high precision
lattice Monte Carlo simulations in  gluodynamics. 
An Abelian gauge field is extracted with no gauge-fixing. 
A static quark-antiquark potential derived from  Abelian Polyakov loop correlators 
gives us the same string tension as the non-Abelian one.  
The Hodge decomposition of the Abelian Polyakov loop
correlator to the regular photon and the singular
monopole parts also reveals that only the monopole
part is responsible for the string tension.
The investigation of the flux-tube profile
then shows that Abelian electric fields defined in
an arbitrary color direction are squeezed by monopole
supercurrents with the same color direction, and
the quantitative features of flux squeezing
are consistent with those observed previously after
Abelian projections with gauge fixing.
Gauge independence of Abelian and monopole dominance strongly
supports that the mechanism of non-Abelian color confinement is
due to the Abelian dual Meissner effect. 
\end{abstract}


\maketitle

Color confinement in  quantum chromodynamics (QCD) 
is still an important unsolved  problem~\cite{CMI:2000mp}. 
't~Hooft~\cite{tHooft:1975pu} and Mandelstam~\cite{Mandelstam:1974pi} conjectured that the QCD vacuum is a kind of a magnetic superconducting state caused by condensation of magnetic monopoles and  an effect dual to the Meissner effect works to confine color charges. 
However, in contrast to SUSY QCD
~\cite{Seiberg:1994rs} or Georgi-Glashow 
model~\cite{'tHooft:1974qc,Polyakov:1976fu} with scalar fields,
to find color magnetic monopoles which
condense is not straightforward in QCD. 

An interesting idea to realize this conjecture is to project 
SU(3) QCD to an Abelian [U(1)]$^2$ theory 
by a partial gauge fixing~\cite{tHooft:1981ht}. 
Then color magnetic monopoles appear as a topological object.
Condensation of the monopoles  causes  the dual Meissner 
effect~\cite{Ezawa:1982bf,Suzuki:1988yq,Maedan:1988yi}.
However there are infinite ways of the above partial gauge-fixing
and whether the 't Hooft scheme is gauge independent or 
not is not clear. Moreover why non-Abelian color charges are confined in the framework of the Abelian mechanism is not clarified.

\par
Numerically, an Abelian projection in non-local 
gauges such as the maximally Abelian (MA)
gauge~\cite{Suzuki:1983cg,Kronfeld:1987ri,Kronfeld:1987vd} 
has been found to support the Abelian confinement scenario
beautifully~\cite{Suzuki:1992rw,Singh:1993jj,Chernodub:1997ay,Bali:1997cp,
Suzuki:1998hc,Koma:2003gq,Koma:2003hv}. Very recently, the present authors have shown 
that the  Abelian dominance and the dual Meissner effect are observed  clearly  also in local unitary gauges such as $F12$ and Polyakov (PL) gauges~\cite{Sekido:2007mp}.  These results strongly suggest that the Abelian confinement mechanism is gauge-independent. 

In this Letter, we study the QCD vacuum after extracting an Abelian link field in a completely gauge-independent way without adopting any special local or non-local gauge fixing. 
We observe that an Abelian confinement mechanism due to condensation of
monopoles is realized. 
A static potential derived from  Abelian Polyakov 
loop correlators gives us the correct string tension.  Moreover only the monopole part in the Abelian Polyakov loop is responsible for the string tension.   
Abelian electric fields defined in an arbitrary 
color direction are squeezed and  the corresponding
monopole currents play the role of   magnetic 
supercurrents. States which are neutral in all color 
directions are not confined and appear as a physical 
state. It is just a color-singlet state. 
Hence, confinement of non-Abelian color charges, 
not that of Abelian charges, is explained in the framework of the gauge-independent Abelian mechanism.  

These findings are completely novel and exciting,
although the continuum and the infinite-volume limits are not studied.

\begin{figure}[bht]
\includegraphics[height=6cm]{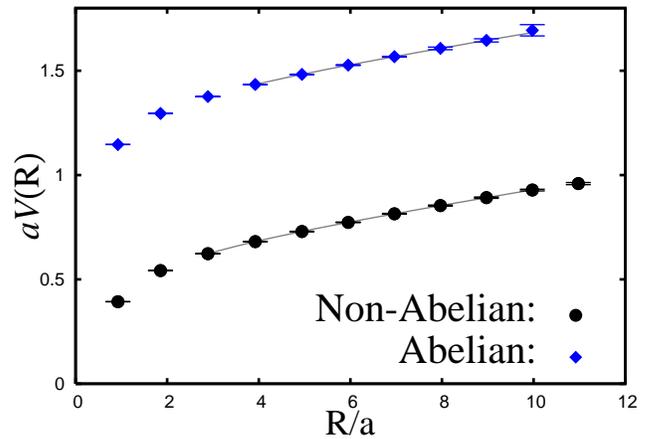}
\vspace{-0.6cm}
\caption{\label{fig-1} The Abelian static potential 
in comparison with the non-Abelian one.
The solid lines denote the best fit to a function $V_{\rm fit}$.}
\end{figure}

Firstly we discuss an Abelian static potential. 
We generate thermalized gluon configurations 
using the  Wilson action 
at a coupling constant~$\beta=2.5$ on the lattice~$N^4=24^4$, 
where the lattice spacing ~$a(\beta)=.0836(8)$~[fm]. 
For simplicity we consider SU(2) gluodynamics,
since essential features are not altered in SU(3).
We extract a $2\times 2$ diagonal Abelian link
field in an arbitrary color direction. 
For example, in the $\sigma_3$ direction, 
\begin{eqnarray*}
U_{\mu}(s) = U^0_{\mu}(s)+i\vec{\sigma}\vec{U}_{\mu}(s)= C_{\mu}(s)
\! \cdot \! \mbox{diag}
\left[e^{i\theta_{\mu}(s)},e^{-i\theta_{\mu}(s)}\right],
\end{eqnarray*}
where 
$\theta_{\mu}(s)=\arctan (U^3_{\mu}(s)/U^0_{\mu}(s))$. 
Note that we can do the same also in
the $\sigma_1$ or $\sigma_2$ direction, 
since all three components are equivalent with no gauge-fixing. 
By using the multi-level noise reduction 
method~\cite{Luscher:2001up}, we
evaluate the Abelian static potential from the 
correlation function of the 
Abelian Polyakov loop operator
\begin{eqnarray}
P_A = \exp[i\sum_{k=0}^{N-1}\theta_4(s+k\hat{4})] \;,
\label{eq-PA}
\end{eqnarray}
separated at a distance $R$.
For the multi-level method, 
the number of sublattices adopted is 6
and the sublattice size is 4.
The results are surprisingly beautiful
as seen from Fig.~\ref{fig-1}.
To reduce the lattice artifact due to finite-lattice cutoff,
we plot the potential using $O(a^2)$ improved distances
~\cite{Necco:2001xg,Luscher:2002qv}.
We try to fit the data to a usual function
$V_{\rm fit}=\sigma R - c/R + \mu$ and find almost 
the same string tension and the Coulombic coefficient
as shown in Table~\ref{stringtension}, indicating 
Abelian dominance.
Here the number of independent vacuum configurations 
is 10 in all cases.
The errors are determined by the jackknife method.
Our results of the string tension are consistent with  theoretical 
observations on the basis of 
reasonable assumptions~\cite{Ogilvie:1998wu,Faber:1998en}.

\begin{table}[tbh]
\begin{center}
\caption{\label{stringtension}Best fitted values of the
string tension $a^2\sigma$,
the Coulombic coefficient $c$ and
the constant $a\mu$.
NA and A-NGF 
denote Non-Abelian and Abelian with no gauge-fixing.  $N_{\rm iup}$
is the number of internal updates in 
the multi-level method. FR means the fitting range.
The $\chi ^2$ for the central value 
is $\chi^2/N_{df} < 0.1$.}
\begin{tabular}{c|c|c|c|c|c}
\hline
& $\sigma a^2$ & $c$ & $\mu a$ & FR(R/a) & $N_{\rm iup}$\\ \hline
NA & 0.0348(7)  & 0.243(6) & 0.607(4) & 3.92 - 9.97 & 15000\\
A-NGF & 0.0352(16) & 0.231(39) & 1.357(17) & 4.94 - 9.97 & 160000
\end{tabular}
\end{center}
\end{table}

\begin{figure}[bht]
\includegraphics[height=5.cm]{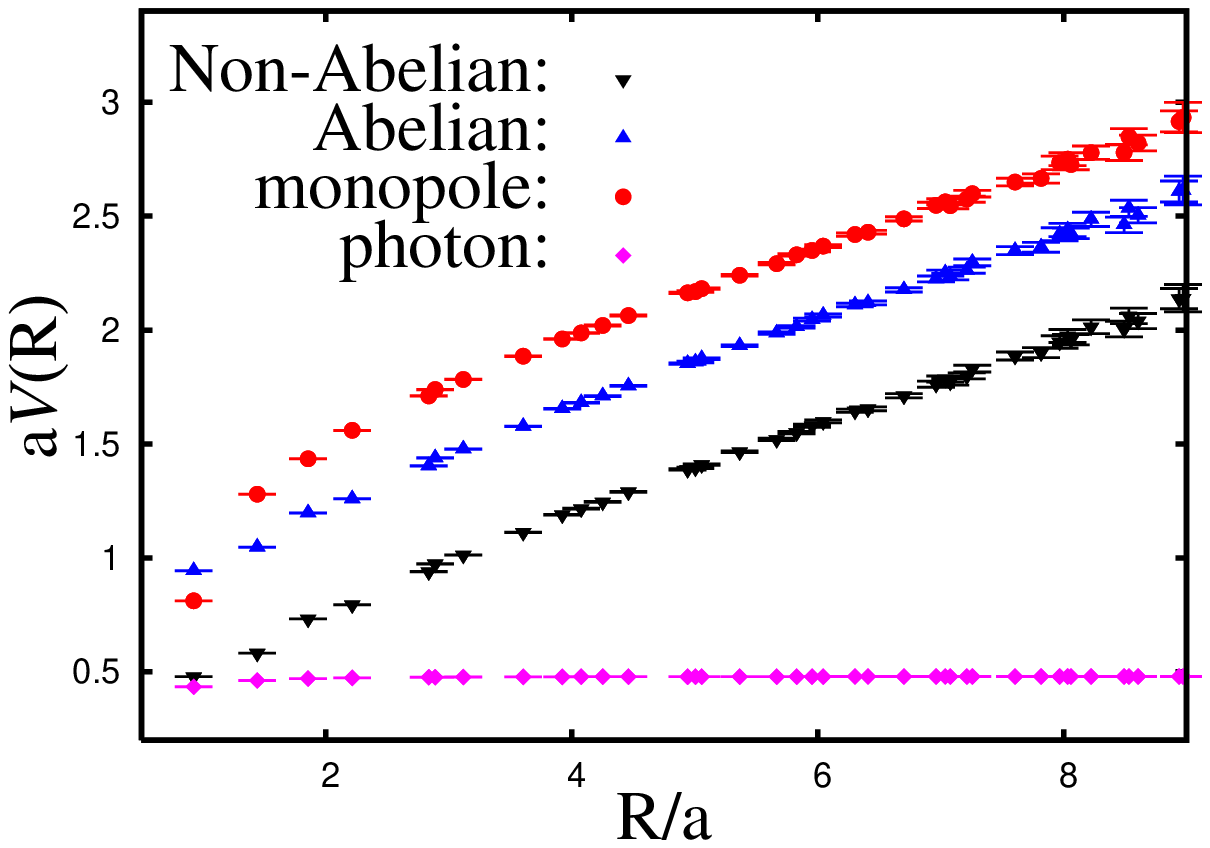}
\includegraphics[height=5.cm]{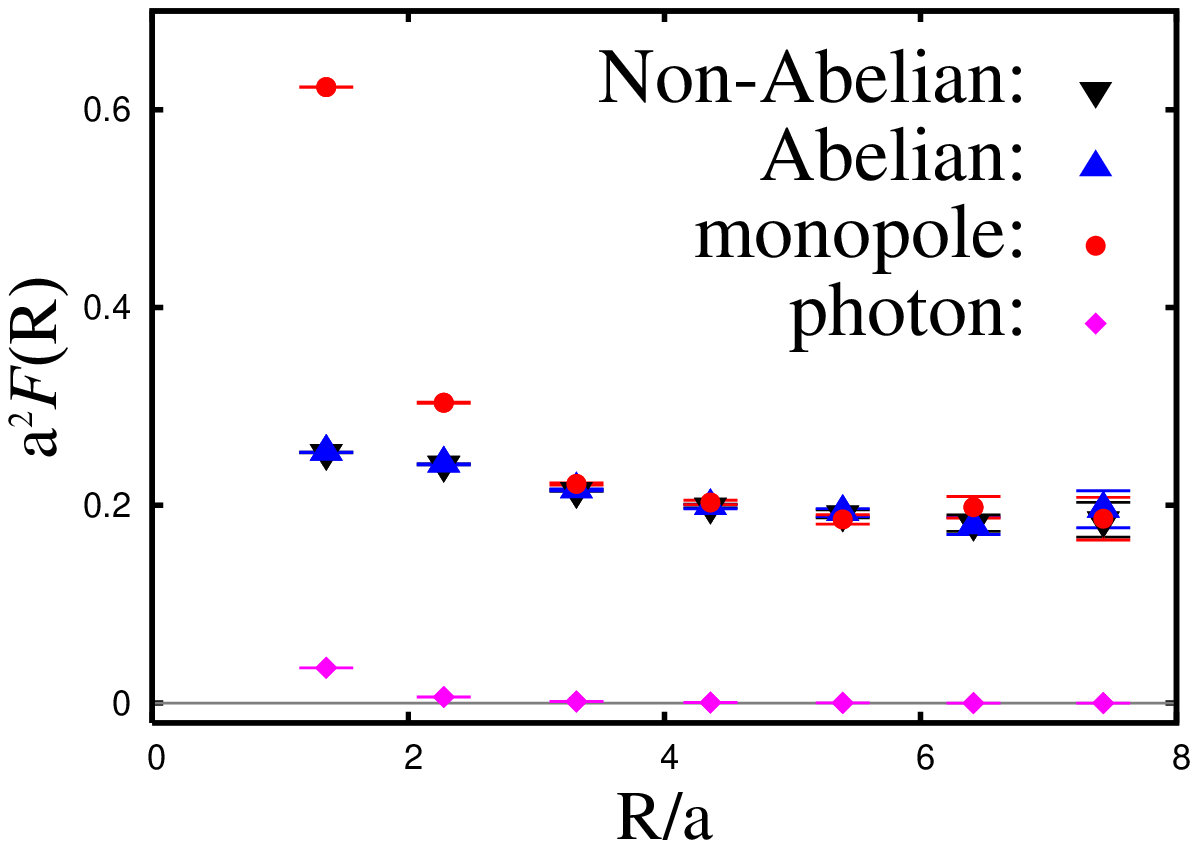}
\vspace{-0.3cm}
\caption{\label{fig-2} 
The static potential (top) and the force (bottom) 
from the Abelian ($P_A$), the monopole ($P_{mon}$)
and the photon contributions ($P_{ph}$) 
in comparison with the non-Abelian ones.
}
\end{figure}

\begin{table}[tbh]
\begin{center}
\caption{\label{stringtension_p}Best 
fitted values of  the
string tension $\sigma a^2$ and the
Coulomb coefficient $c$.
M-NGF (P-NGF) 
denotes the monopole (the photon) part.}
\begin{tabular}{c|c|c|c|c|c}
\hline
& $\sigma a^2$ & $c$ & $\mu a$ & FR(R/a) & $\chi^2/N_{df}$ \\ \hline
 NA   & 0.181(8)  & 0.25(15) & 0.54(7)  & 3.92 - 8.50 & 1.00 \\ 
A-NGF & 0.183(8)  & 0.20(15) & 0.98(7)  & 3.92 - 8.23 & 1.00 \\ 
M-NGF & 0.183(6)  & 0.25(11) & 1.31(5)  & 3.92 - 6.71 & 0.98 \\ 
P-NGF &-0.0002(1) & 0.010(1) & 0.48(1)  & 4.94 - 9.44 & 1.02
\end{tabular}
\end{center}
\end{table}

Secondly we discuss the role of monopole contribution. 
The monopole part of the operator can be extracted as follows.
The Abelian Polyakov loop~\eqref{eq-PA} can be written 
by a product of a
photon and a Dirac-string parts~\cite{Suzuki:1994ay}.
Note that
\begin{equation}
\theta_4 (s)= -\sum_{s'} D(s-s')[\partial'_{\nu}\theta_{\nu 4}(s')+
\partial_4 (\partial'_{\nu}\theta_{\nu}(s'))] \;, \label{t4}
\end{equation} 
where $D(s-s')$ is the lattice Coulomb propagator, $\theta_{\mu\nu}(s)=\partial_{\mu}\theta_{\nu}(s)-\partial_{\nu}\theta_{\mu}(s)$ 
and $\partial_{\nu}(\partial'_{\nu})$ is a forward(backward) difference. We have used 
$\partial_{\nu}\partial'_{\nu}D(s-s')=-\delta_{ss'}$.
The second term in the right-hand side of 
(\ref{t4}) does not contribute to the Abelian Polyakov loop (\ref{eq-PA}). Now
$\theta_{\mu\nu}(s) =\bar{\theta}_{\mu\nu}(s)+2\pi 
n_{\mu\nu}(s)\ \ (|\bar{\theta}_{\mu\nu}|<\pi)$, 
where $n_{\mu\nu}(s)$ is an integer 
corresponding to the number of the Dirac string.
Hence we obtain $P_A=P_{ph} \cdot P_{mon}$, where
\begin{eqnarray*}
P_{ph} & = &\exp\{-i\sum_{k=0}^{N-1} \!\sum_{s'}
D(s+k\hat4-s')\partial'_{\nu}\bar{\theta}_{\nu 4}(s')\},\\
P_{mon} & = &\exp\{-2\pi i\sum_{k=0}^{N-1}\! \sum_{s'}
D(s+k\hat4-s')\partial'_{\nu}n_{\nu 4}(s')\}.
\end{eqnarray*}
 We call $P_{ph}$ and $P_{mon}$ the photon and 
the monopole contributions, respectively, 
since the Dirac string $n_{\beta\gamma}(s)$ leads 
us to a monopole current
$k_{\mu}(s)= (1/2)\epsilon_{\mu\alpha\beta\gamma}\partial_{\alpha}
n_{\beta\gamma}(s+\hat\mu)$~\cite{DeGrand:1980eq}.

We need a
non-local Coulomb propagator in the separation, 
so that the multi-level
noise reduction method cannot be applied in this case. 
Here we consider a $T\neq 0$ system in the confinement
phase  with the Wilson action on $24^3\times 4$ lattice. We use
 about 6000 thermalized configurations  
at $\beta=2.2$, where
the lattice spacing is $a(\beta)= .191(8)$~[fm]. 
Since the expectation values of 
the correlation functions of
$P_A$, $P_{ph}$ and $P_{mon}$ are still very small with 
no gauge-fixing,
we adopt a new noise reduction method.
For a thermalized vacuum ensemble,
we produce many gauge copies applying random gauge transformations,
compute the operator for each copy, and 
take the average of all copies.
Note that as long as a gauge-invariant operator is evaluated, 
such copies are identical, 
but they are not if a gauge-variant 
operator is evaluated.
Practically, we prepare $1000$ gauge copies for each 
configuration. 
We also apply one-step hypercubic
blocking (HYP)~\cite{Hasenfratz:2001hp}
to the temporal links for further noise reduction.

We obtain very good signals
for the Abelian, the monopole and the photon 
contributions to the static potential 
as shown in Fig.~\ref{fig-2}. 
We try to fit the potential in Fig.~\ref{fig-2} 
to the function $V_{\rm fit}$ 
and extract the string tension and the Coulombic 
coefficient of each potential as summarized 
in Table~\ref{stringtension_p}. 
Abelian dominance is seen again beautifully in this case. 
Moreover, we can see monopole dominance, namely, only
the monopole part of the Polyakov loop correlator is
responsible for the string tension. The photon part has no
linear potential. The agreement 
among the string tensions coming from
non-Abelian, Abelian and monopole Polyakov 
loop correlators is almost perfect as seen also from the force in Fig.~\ref{fig-2}
in comparison with
the MA case, where only 80-90 percent agreement is
observed at finite lattice spacings.
The short-range behavior of the potential may be 
affected by HYP.

Thirdly we discuss
the Abelian dual Meissner effect. 
We investigate  the Abelian
flux-tube profile by
evaluating connected correlation
functions~\cite{Cea:1995zt, DiGiacomo:1989yp}
between a Wilson loop $W$ and 
Abelian operators ${\cal O}_A$
constructed by Abelian link fields,
\begin{eqnarray*}
\langle {\cal O}_A(r) \rangle_{W}&=&
{\langle \mbox{Tr}\left[
LW(r=0,R,T)L^{\dagger}\sigma^3{\cal O}_A(r)
\right]\rangle
\over \langle \mbox{Tr}\left[W(R,T)\right]\rangle},
\end{eqnarray*}
where $L$ is a product of non-Abelian link fields 
(a Schwinger line) connecting the Wilson loop
with the Abelian operator.
We may use the cylindrical coordinate
$(r,\phi,z)$ to parametrize the
the $q$-$\bar{q}$ system, where the $z$ axis
corresponds to the $q$-$\bar{q}$ axis and
$r$ to the transverse distance.
We are interested in the field profile as a function 
of $r$ on the mid-plane of the $q$-$\bar{q}$ distance.
In this calculation, we employ the improved Iwasaki gauge
action~\cite{Iwasaki:1985we} with the coupling constant
$\beta= 1.20$,
which corresponds to the lattice spacing
$a(\beta)= .0792(2)$~[fm]~\cite{Suzuki:2004dw}.
The lattice volume is $32^4$ with periodic boundary conditions.
We generate 4000 thermalized configurations.
To improve a signal-to-noise ratio,
the APE smearing technique is
applied to the Wilson loop~\cite{Albanese:1987ds}.

\begin{figure}[t]
\includegraphics[height=5.cm]{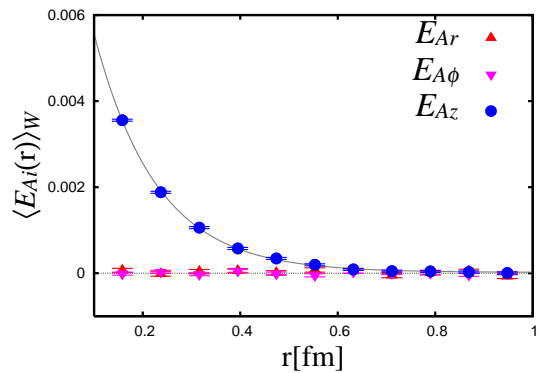}
\vspace{-0.4cm}
\caption{\label{fig-3}
The profile of the Abelian electric fields
for $W(R=5a,T=5a)$.}
 \vspace{-0.3cm}
\end{figure}

\begin{figure}[t]
\includegraphics[height=5.cm]{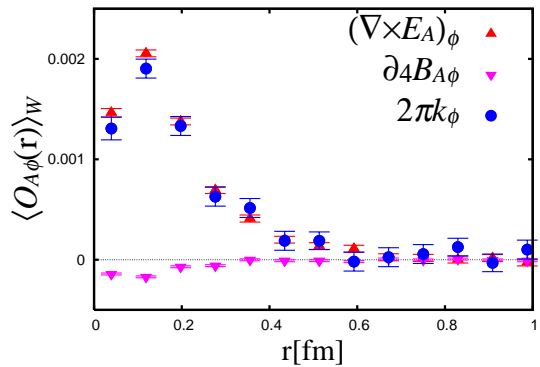}
\vspace{-0.4cm}
\caption{\label{fig-4} The curl of the Abelian electric field, 
magnetic displacement currents and monopole currents 
for $W(R=5a,T=5a)$.}
 \vspace{-0.3cm}
\end{figure}

We measure all components of the Abelian electric fields 
$E_{Ai}(s)=\bar{\theta}_{4i}(s)$ and find that
only $E_{Az}$ is squeezed as shown in Fig.~\ref{fig-3}.
We try to fit $\langle E_{Az} \rangle_{W}$ to
a function $f(r)=c_1\exp(-r/\lambda)+c_0$.
Here $\lambda$ corresponds to the penetration length. 
We obtain $\lambda=0.128(2)$~[fm], 
which is similar to those obtained in the MA 
gauge and unitary gauges~\cite{Sekido:2007mp} as seen from Table \ref{penetration}.

\begin{figure}[thb]
\includegraphics[height=5.cm]{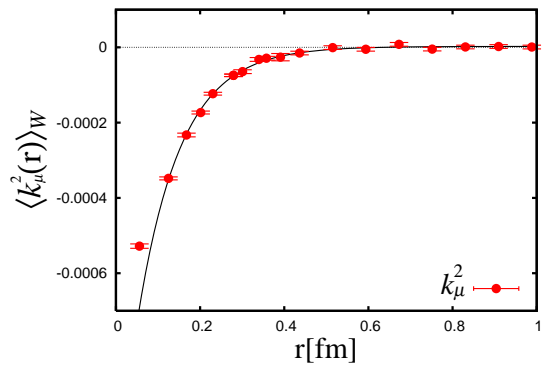}
\vspace{-0.4cm}
\caption{\label{fig-5} The correlation between the
Wilson loop and the
 squared monopole density for $W(R=5a,T=5a)$. 
The solid line denotes the best exponential fit.}
\vspace{-0.3cm}
\end{figure}
\if01
\begin{table}[t]
\caption{\label{penetration}
The penetration lengths $\lambda$~[fm],
the coherence lengths $\xi$~[fm] and the GL parameter
 $\kappa=\lambda/\xi$ for $W(R=5a,T=5a)$. 
 NGF, MA and PL denote no gauge fixing, MA gauge
 and Polyakov gauge. The data of the MA and PL gauges
 are taken from Ref.~\cite{Sekido:2007mp}.}
\begin{tabular}{c|ccc}
&NGF & MA & PL \\
\hline
 $\lambda$ 
 &0.128(2)&0.129(2) & 0.132(4) \\
 $\xi/\sqrt{2}$ &0.102(3)& 0.109(4)  & 0.101(20) \\
\hline
 $\sqrt{2}\kappa$ & 1.25(6) & 1.18(6) & 1.31(29)
\end{tabular}
\vspace{-0.3cm}
\end{table}
\fi

To see what squeezes the Abelian electric field, 
let us study the Abelian (dual) Amp\`ere law 
\begin{eqnarray*}
\vec{\nabla}\times\vec{E}_{A}=
\partial_{4}\vec{B}_{A}+2\pi\vec{k}\;, 
\end{eqnarray*}
where $B_{Ai}(s)=(1/2)\epsilon_{ijk}\bar{\theta}_{jk}(s)$.
Each term is evaluated on the same mid-plane
as for the electric field.
We find that only the azimuthal components are non-vanishing,
which are plotted in Fig.~\ref{fig-4}.
Note that if the electric field is purely of the Coulomb 
type, the curl of electric field is zero. 
Contrary, the curl of the electric field
is non-vanishing and is reproduced only by monopole currents. 
The magnetic displacement current is almost vanishing. 
These behaviors are clearly a signal of the Abelian dual
Meissner effect, which are quite the same as those 
observed in the MA gauge~\cite{Koma:2003gq,Koma:2003hv}. 

Fourthly, we may estimate the vacuum type by  evaluating also
the coherence length~$\xi$ from the correlation function between
the Wilson loop and the squared monopole density 
$k_{\mu}^2(s)$
~\cite{Chernodub:2005gz}.
The correlation function is plotted in Fig.~\ref{fig-5}
and the coherence length extracted from the functional form
$g(r)=c_{1}'\exp (-\sqrt{2}r/\xi) + c_{0}'$
is $\xi/\sqrt{2}=0.102(3)$~[fm].  
The GL parameter $\sqrt{2}\kappa=\lambda/\xi=1.25(6)$ is 
close to the values obtained with gauge fixing\cite{Sekido:2007mp}.
Since the Wilson loop used here may still be small, 
what we can say is that the vacuum type is near 
the border between the type~1 and~2.

To summarize, we have observed gauge-independence of 
Abelian and monopole dominance for the string tension
and of the Abelian dual Meissner effect 
in gluodynamics by using lattice Monte Carlo simulations.
These results are quite remarkable in the sense that 
confinement of non-Abelian color charges can be explained
in the framework of the Abelian dual Meissner effect. 
Since no gauge-fixing is done, gauge fields  in any 
color direction are equivalent. Abelian electric fields
in all color directions are squeezed due to monopoles in
the corresponding color direction.   An Abelian neutral
state in all color directions can survive as a physical 
state, and such a state is only the color singlet state.
For example, consider meson states $u_c\bar{u}_c$ and
$d_c\bar{d}_c$, where $u_c ~(d_c)$ is an eigenstate
of $\sigma_3/2$ with an eigenvalue $1/2~(-1/2)$. 
These are Abelian neutral in the $\sigma_3/2$ direction.
Similarly, $U_c\bar{U}_c$ and $D_c\bar{D}_c$ are Abelian
neutral in the $\sigma_1/2$ direction, where 
$U_c=(u_c+d_c)/\sqrt{2} ~(D_c=(u_c-d_c)/\sqrt{2})$ is an
eigenstate of $\sigma_1/2$. Note that $u_c\bar{u}_c$
($U_c\bar{U}_c$) and $d_c\bar{d}_c$ ($D_c\bar{D}_c$) contain
both Abelian charged and neutral states in the $\sigma_1/2$
($\sigma_3/2$) direction. But a SU(2) singlet state 
$u_c\bar{u}_c+d_c\bar{d}_c=U_c\bar{U}_c+D_c\bar{D}_c$
is Abelian neutral in all color directions.
Hence confinement of non-Abelian color charge can be 
explained in terms of the Abelian confinement scenario 
of the dual Meissner effect.
  
Finally it is interesting to study the relation between
the violation of the non-Abelian Bianchi
identity~\cite{Skala:1996ar} and Abelian monopoles 
with no gauge-fixing. 
  
The authors are grateful to M. Polikarpov, V. Zakharov, M. Chernodub, V. Bornyakov and G. Schierholz for useful discussions. The numerical simulations of this work were done using RSCC computer clusters in 
RIKEN, SX5 and SX8 at RCNP of Osaka University 
and SX8 at YITP in Kyoto University. The authors would like to thank RIKEN, RCNP and YITP for their support of computer facilities. 


\end{document}